**Enter: Graduated Realism: A Pedagogical Framework for AI-Powered Avatars in Virtual Reality Teacher Training**

by Judson Leroy Dean Haynes IV          jhaynes2@umd.edu          University of Maryland

**Abstract**

Virtual Reality (VR) simulators are emerging as a powerful tool to provide pre-service teachers with essential classroom practice. The integration of generative Artificial Intelligence (AI) to power student avatars is a significant technological leap, but it introduces a critical new challenge: determining the optimal level of avatar realism for effective pedagogical outcomes. This literature review critically examines the evolution of avatar realism in VR teacher training, synthesizes its theoretical implications, and proposes a new pedagogical framework to guide future design and implementation. Through a systematic review, this paper traces the progression from early human-controlled avatars to current generative AI-driven prototypes. Established learning theories, such as Cognitive Load Theory and the Cognitive Affective Model of Immersive Learning (CAMIL), are applied to argue that hyper-realism is not always optimal. High-fidelity avatars can impose excessive extraneous cognitive load on novice teachers, an idea supported by recent empirical findings. A significant gap is identified between the technological drive for photorealism and the pedagogical need for scaffolded learning. To address this gap, the paper proposes Graduated Realism, a framework that advocates for starting trainees with lower-fidelity avatars and progressively increasing behavioral complexity as skills develop. To make this framework computationally feasible, the paper outlines a novel single-call architecture, termed "Crazy Slots", which uses a probabilistic engine and a rich Retrieval-Augmented Generation (RAG) database to generate authentic, real-time student responses without the latency and cost of multi-step reasoning models. This review provides evidence-based principles for designing the next generation of AI simulators, arguing that a pedagogically grounded approach to realism is essential for transforming these tools into scalable and truly effective components of teacher education.

**Introduction**

Teacher education programs have long grappled with the challenge of providing sufficient practical experience to pre-service teachers before they enter real classrooms. Virtual Reality (VR) simulators have emerged as a promising solution, offering immersive, risk-free environments where teachers can practice and hone their skills, representing a significant trend in the convergence of AI and immersive technologies in education (Lampropoulos, 2025). Early VR-based teacher training systems demonstrated that virtual "classrooms" with student avatars could bridge the gap between educational theory and practice, giving novices a safe space to apply pedagogical concepts and receive feedback. However, these initial systems often required significant human labor, with role-players needing to "puppeteer" student avatars in real time. Recent advances in Artificial Intelligence (AI), especially large language models (LLMs) like OpenAI's ChatGPT, are transforming this landscape. By integrating AI-driven student avatars into VR simulations, developers can create more responsive and potentially more realistic classroom scenarios without the constant need for human operators.

This technological shift places a new and critical focus on the realism of the AI student avatars themselves. This pursuit is not without peril, as it forces developers to navigate the so-called 'uncanny valley,' where avatars that are almost, but not quite, fully human-like can evoke unease and hinder the learning process (Byrne, 2025; Mori, 1970). Realism in this context, therefore, extends beyond mere visual fidelity; it encompasses behavioral authenticity, interactional responsiveness, and crucially, appropriate cultural and individual representation. While the drive towards greater realism seems intuitive for mimicking authentic classroom experiences, it raises vital pedagogical and ethical questions. Is photorealism always optimal, particularly for novice teachers? Might a graduated realism approach—perhaps starting trainees with less visually complex, stylized avatars (e.g., 'anime style' with clear expressions) before progressing to higher fidelity—be more effective for scaffolding skills and managing anxiety? This question is gaining empirical traction, with recent studies beginning to explore how a user's awareness of an avatar's AI nature can impact learning outcomes and trust, suggesting that the psychological effect of realism is a critical factor (Westin & Öberg, 2024).

This literature review critically examines the evolution of student avatar realism within VR teacher training simulators. It explores the theoretical implications of varying realism levels, scrutinizes the technical challenges and significant ethical considerations in designing realistic and representative avatars, and identifies a critical gap concerning the pedagogical effectiveness of different realism strategies, particularly graduated realism.

### Research Questions

1. **RQ1 – Mapping the field:** What ranges of avatar realism (visual, behavioral, interactional) have been operationalized in VR teacher-training studies to date?
2. **RQ2 – Learning impact:** How does the degree of avatar realism influence preservice teachers' cognitive load, perceived social presence, and transfer of classroom-management skills?
3. **RQ3 – Design guidance:** What evidence-based principles can be synthesized into a *graduated realism* framework that aligns avatar fidelity with specific instructional objectives and resource constraints?

## Methodology

This literature review employed a systematic approach to identify and synthesize relevant scholarly work concerning the realism of AI-powered student avatars within virtual reality (VR) teacher training simulations. The primary goal was to understand the evolution, theoretical implications, pedagogical considerations, and ethical challenges associated with varying levels of avatar realism in this specific context.

## Search Strategy

The literature search was conducted using major academic databases, including Google Scholar, Scopus, ERIC, PsycINFO, and the ACM Digital Library. Search strategies combined keywords related to the core components of the review. Primary search strings included combinations and variations of terms such as:

- ("AI" OR "artificial intelligence" OR "LLM" OR "large language model") AND ("avatar" OR "virtual student" OR "virtual agent") AND ("virtual reality" OR "VR" OR "immersive simulation") AND ("teacher education" OR "teacher training" OR "pre-service teacher")
- Additional searches incorporated terms related to realism: ("avatar realism" OR "fidelity" OR "authenticity" OR "presence" OR "immersion") AND ("teacher simulation" OR "VR training") AND ("simulation-based+teacher", or "teaching simulation")
- Keywords derived from seminal works identified early in the process (e.g., specific simulator names like TeachLivE, Mursion; key author names) were also used for targeted searches.
- Traditional database searches were supplemented using AI-driven academic search tools to identify potentially related concepts and relevant publications, which were subsequently verified and screened using the criteria below.

## Inclusion and Exclusion Criteria

To ensure relevance and rigor, the following criteria were applied:

- **Inclusion Criteria:**
  - Studies explicitly discussing the use of VR simulations for teacher education or training.
  - Simulations involving student avatars, particularly those driven or augmented by AI.
  - Discussion relevant to avatar realism (visual, behavioral, interactional), pedagogical implications, theoretical underpinnings, or ethical considerations.
  - Peer-reviewed journal articles, conference proceedings, and relevant scholarly books or chapters.
  - Publications primarily in English.
  - Focus on recent advancements, generally considering literature published from approximately 2010 onwards to capture the rise of modern VR and AI integration, with particular emphasis on work from the last 5-7 years reflecting the impact of LLMs.
- **Exclusion Criteria:**
  - Studies focusing solely on non-immersive simulations without a VR component.
  - VR applications outside the context of teacher education/training.
  - Simulations using avatars solely controlled by human operators ("puppeteering") without AI elements, unless discussed for historical context.
  - Purely technical papers lacking discussion of pedagogical, theoretical, or ethical aspects.
  - Editorials, opinions, or non-scholarly articles.

## Screening and Selection Process

Initial search results were screened based on titles and abstracts to remove duplicates and obviously irrelevant studies. Potentially relevant articles underwent a full-text review to determine final eligibility based on the inclusion/exclusion criteria. The reference lists of key articles were also scanned to identify additional relevant sources (snowballing). This iterative process resulted in the body of literature analyzed in this review.

**Roadmap for the Review**

To structure this analysis, the review will proceed as follows:

1. **The Evolution of Student Avatar Realism:** Tracing the progression from human-controlled avatars (like those in TeachLivE) to current generative AI-powered prototypes (exemplified by studies like Docter et al., 2024), focusing specifically on advancements and limitations in achieving visual, behavioral, and interactional realism.
2. **Theoretical Implications of Avatar Realism:** Analyzing how different levels of realism might impact teacher learning through the lens of established theories, including immersive learning models (e.g., CAMIL), cognitive load theory, social learning theory, and concepts of psychological safety.
3. **The Pedagogical Case for Graduated Realism:** Elaborating on the hypothesis that starting with lower-fidelity avatars and progressing to higher fidelity may be a beneficial scaffolding strategy, addressing potential impacts on anxiety, cognitive load, and skill acquisition.
4. **Ethical and Cultural Challenges in Avatar Design:** Discussing the critical issues surrounding bias, representation, data privacy (especially if using real student data for training), and the complexities of simulating realistic but potentially sensitive student behaviors.
5. **Gaps, Limitations, and Future Directions:** Synthesizing the identified gaps in current research. While nascent empirical work is beginning to investigate the impact of avatar awareness on cognitive load and trust (Westin & Öberg, 2024), a significant gap remains in studies that systematically test a progressive, scaffolded model of graduated realism.
6. **Conclusion:** Summarizing the findings and reiterating the importance of thoughtfully navigating the complexities of avatar realism to harness the full potential of AI-driven VR simulators for teacher education.

**The Evolution of Student Avatar Realism**

The initial generation of VR or mixed-reality teacher training environments, exemplified by systems like TeachLivE™ developed in the early 2010s, relied heavily on human operators to animate student avatars. In these platforms, trained "interactors" (or puppeteers) controlled multiple avatars in real-time, using tools like voice changers to simulate different student personalities and dialogue (Ersozlu et al., 2021). Consequently, the behavioral and conversational realism of the avatars was entirely dependent on the skill, training, and moment-to-moment performance of the human operator.

While such simulations proved useful for practicing specific teaching skills (Ersozlu et al., 2021), this reliance on human control presented significant limitations. These systems were often resource-intensive, requiring skilled operators and complex scheduling, and sometimes offered limited visual immersion on standard 2D screens (Huang et al., 2022). Critically, this human-in-the-loop model also impacted the experiential realism for the trainee. For example, studies on these early immersive systems noted that trainees struggled with the simulation's authenticity because the avatars had fixed facial expressions, limited interactivity, and an inability to handle open-ended questions, which hindered the feeling of a natural classroom interaction (Chen, 2022; van der Zee et al., 2025). These combined challenges related to cost, accessibility, and the inherent variability of human-driven realism spurred the move towards integrating Artificial Intelligence (AI) to automate avatar control and explore new possibilities for enhancing realism and responsiveness.

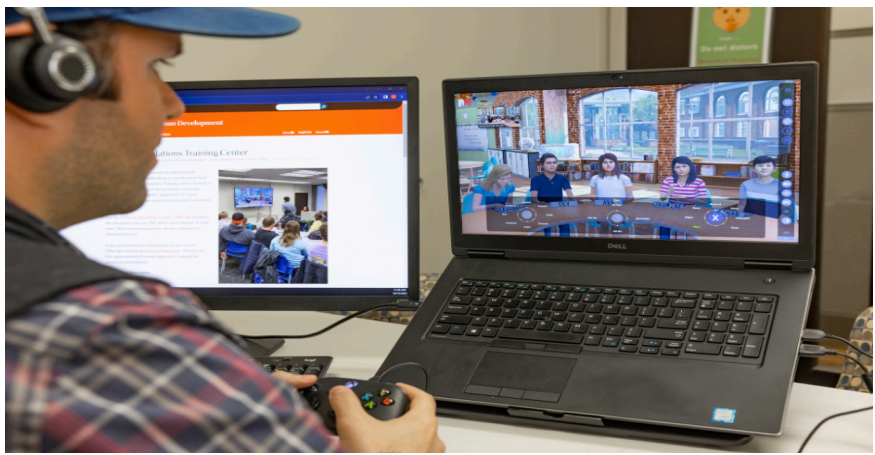

**Figure 1.** A pre-service teacher engages with a virtual classroom simulation (Bowling Green State University, 2023)

## Early AI Integration

A commercial successor, Mursion, centralized the puppeteer approach by providing remote trained interactors that institutions could schedule on demand (Yoodli, 2024). This made the system more scalable (even extending beyond K-12 classrooms to corporate training scenarios ), but it retained the fundamental dependence on human-in-the-loop control (Yoodli, 2024).

 A key milestone in the evolution of these simulators was the gradual incorporation of artificial intelligence to reduce reliance on human controllers. Even Mursion includes a modest AI element: the student avatars' non-verbal behaviors (gestures, posture, and facial expressions) were partly controlled by an AI system, reflecting the state of AI at the time, making their movements more natural without direct puppeteer input (Yoodli, 2024). This was one of the first instances of AI in classroom simulators, though the conversational and decision-making aspects still relied on humans. Over time, advances in AI–especially in natural language processing–opened the door to more autonomous virtual students. Developers began experimenting with "semi-autonomous agents" that could handle certain student behaviors automatically. For example, research prototypes allowed an instructor to adjust parameters (like a student agent's level of off-task behavior or responsiveness) and the system would generate corresponding avatar actions from a predefined range (Seufert et al., 2022). These early

AI-driven agents were limited (often using scripted or menu-based responses), and technical bugs sometimes disrupted the illusion of a smooth classroom interaction (Mouw et al., 2020). Nevertheless, they hinted at the potential for fully AI-driven student avatars that could converse and react in real-time, allowing more flexible, unscripted practice.

**The Leap with Generative AI/LLMs**

The most recent leap in VR teacher training simulators involves integrating generative AI LLMs to control student avatar dialogue and behavior, exemplified by Docter et al. (2024) state-of-the-art proof-of-concept. Their system uses the generative power of ChatGPT to control multiple student avatars in an immersive VR classroom, meaning an AI generates the students' verbal responses and some behaviors on the fly, replacing the need for a human puppeteer (Docter et al., 2024). Trainees in a VR headset interact with these avatars, which dynamically respond to instructions or missteps based on the AI (Docter et al., 2024). To model diverse reactions, Docter et al. implemented an algorithmically controlled 'mood factor' for each avatar, shifting based on trainee actions (analyzed via GPT-3.5) and contextual factors like peer influence or time (Docter et al., 2024). This allows avatars to realistically simulate behaviors like engaging in disruptive conversations or losing focus, creating dynamic scenarios (Docter et al., 2024). By leveraging generative AI, this innovation effectively eliminates the human bottleneck of earlier systems, allowing simulators to present a potentially infinite range of unscripted scenarios and dialogues. This approach greatly increases training flexibility—the same platform can simulate vastly different classroom atmospheres simply by tweaking AI parameters—and significantly reduces instructor workload.

Beyond the Docter et al. study, other projects are also blending VR with advanced AI. Researchers at the University of Michigan, for example, created a system for cultural heritage education where an AI-driven tutor avatar in VR engages users in natural language conversation (Gao et al., 2024). While not focused on teacher training, this work shows the broader trend: AI avatars can now listen to spoken input, use an LLM to generate contextually relevant replies, and speak back with a synthetic voice (Fink et al., 2024). The open-source project GPTAvatar is another state-of-the-art AI avatar framework that records a user's speech, converts it to text, feeds it to an LLM (like GPT-3/4) to generate a response, and then uses text-to-speech to voice the avatar's reply (Fink et al., 2024). GPTAvatar, built on the Unity game engine, even allows setting an avatar's personality and scenario, illustrating how customizable and powerful AI-driven avatars have become (Fink et al., 2024).

Another breakthrough advancement in LLM-driven simulation comes from Wu et al. (2025). In their study, titled 'Embracing Imperfection: Simulating Students with Diverse Cognitive Levels Using LLM-based Agents,' the team utilized a Python programming student dataset to construct customized student knowledge graphs, which served as dynamic cognitive prototypes (Wu et al., 2025). These prototypes were then referenced by multiple LLM calls to simulate students with diverse cognitive levels; to ensure the simulated student answers precisely reflected their unique imperfections and knowledge gaps, the framework employed a multi-stage self-refinement process (Wu et al., 2025). This involved iteratively generating and evaluating several possible answers (using a 'beam search' approach) through multiple LLM calls, progressively improving the fidelity of the simulated student's response until it aligned with their specific cognitive profile (Wu et al., 2025). Wu et al. (2025) successfully addressed the issue of AI's tendency to always provide correct answers, demonstrating a method for consistent, plausible imperfection. Their approach consistently outperformed baseline models, achieving a "100% improvement in simulation accuracy" (Wu et al., 2025, p. 1). In a similar vein, Tonga et al. (2025) present a large-scale simulation of tutor-student interactions using LLMs, where a weaker model simulates the student, and a stronger model plays the tutor role. Their work investigates the impact of multilingual feedback on student learning outcomes in math, further demonstrating the growing sophistication of AI in modeling student behavior (Tonga et al., 2025).

These technological advances suggest that VR teacher simulators can be populated with multiple AI students to simulate entire classrooms with diverse student personalities. As LLMs improve on their multimodal capabilities (visual and auditory), avatar realism will be enhanced both behaviorally and interactionally. They will be able to "see" the virtual environment and the teacher's actions. For example, an AI student could observe where the teacher writes on the virtual board or note the teacher's physical proximity, and adjust its behavior accordingly. Early signs of this are already evident: the latest GPT-4 model can interpret images and "camera feed" with advanced computer vision, meaning an AI-driven avatar could respond to visual cues, not just spoken words (Fink et al., 2024). This paves the way for truly lifelike simulations, where virtual students react to a teacher moving around the room, pointing at objects, or displaying instructional materials. However, a main issue is that results for multimodal capabilities are very slow and expensive to output (Wang et al., 2024) (anything AI generated besides dialogue), and we'll have to wait for these processes to quicken.  In short, the evolution of VR teacher training tools has moved from human-operated avatars to increasingly autonomous AI-driven students. This evolution promises greater realism, scalability, and variety in teacher practice scenarios than ever before.

**Realism**

Defining 'realism' for VR student avatars requires considering multiple dimensions beyond simple appearance. It encompasses visual fidelity (the avatar's look, from stylized to photorealistic), behavioral authenticity (believable dialogue, non-verbal cues, consistent personality, and appropriate student-like actions, including potential misbehavior), and interactional realism (responsive engagement with the trainee, environment, and context, ideally with continuity). Technologically, creating highly realistic environments and avatars approaching photorealism is becoming increasingly feasible, as exemplified by projects aiming for high-fidelity simulations like TaoAvatar (Chen et al., 2025).

While achieving near-indistinguishable realism might seem like an ultimate goal for simulation fidelity, its suitability as a starting point for novice teacher training is questionable. The powerful sense of immersion achievable in VR doesn't solely depend on photorealistic characters; experiences like the 'Richie's Plank Experience' game (Meta, n.d.) illustrate how compelling environmental design and perceived physical risk can create strong presence and elicit real-world reactions, even with simpler visuals. This suggests factors contributing to interactional realism and immersion can be highly effective independent of extreme visual fidelity in the avatars themselves.

Furthermore, striving for near-perfect visual fidelity carries the specific risk of encountering the 'uncanny valley'—the phenomenon where characters that are almost, but not quite, fully humanlike can evoke unease or eeriness (Mori, 1970; Kätsyri et al., 2015). However, a potentially broader concern for novice training is that interacting with highly realistic avatars portraying complex or challenging behaviors might be overly intimidating, increase performance anxiety, or impose excessive cognitive load compared to interacting with more stylized representations. This concern is particularly salient when considering sophisticated behavioral models like those developed by Wu et al. (2025), which aim to accurately simulate students with nuanced cognitive levels and imperfections, potentially requiring complex interpretation from trainees. Similar to how a cartoon depiction might lessen the perceived intensity of an action compared to a photorealistic one, stylized avatars with clear expressions could provide a less overwhelming environment for beginners.Therefore, this review proposes investigating a graduated realism approach.

## Conceptual Learning Theories in Avatar-Based Training

The development and research of VR teacher simulators are grounded in several conceptual learning theories and frameworks that explain why these environments can be effective. A consistent theme in the literature is the blending of theory and practice—using VR to make abstract pedagogical theories more concrete for teachers in training (Dieker et al., 2014; Huang et al., 2022). For instance, McGarr (2020) observes that virtual classroom simulations help bridge the gap between theoretical knowledge from coursework and the practical skills needed in real classrooms by providing "vivid, example-based feedback" in a safe setting. In a VR simulation, a pre-service teacher can attempt to apply a strategy (say, a behavior management technique learned in class) and immediately see its effect on student avatars, which reinforces conceptual learning with experiential feedback. This iterative practice aligns with Kolb's (1984) experiential learning cycle (concrete experience → reflective observation → abstract conceptualization → active experimentation), effectively turning teaching theory into lived experience for the trainee. VR simulations also lower performance anxiety for novice teachers, as they can practice classroom management or instructional techniques without the fear of harming real students or being judged by supervisors (McGarr, 2020). This safe environment encourages experimentation and learning from mistakes, an idea rooted in constructivist theory – learners (in this case, teachers) construct knowledge through experience and reflection, especially when they can repeatedly practice and refine their approach.

## Zone of Proximal Development

Several studies emphasize the importance of scaffolding and feedback in VR learning environments (Dieker et al. 2014; Fink et al., 2024). Scaffolding, based on Vygotsky's (1978) concept of the Zone of Proximal Development (ZPD), involves providing support to learners as they develop new skills. In VR teacher training, scaffolding can be implemented via programmed coaching or guided reflection between teaching simulations (Ke et al. 2020). For example, a simulator might allow a mentor (or an AI coach) to pause the action and give feedback, or it might have levels of difficulty that gradually increase as the teacher gains proficiency. Dieker et al. (2014) describe the teaching simulation process as a "cyclical process of feedback," wherein teachers can attempt a lesson, witness the outcome, receive feedback, and then try again, iteratively improving their practice. This cyclic feedback loop is a practical application of reflective practice theory (Schön, 1983), enabling teachers to reflect-in-action and reflect-on-action in a controlled setting. By structuring VR training like a video game with progressive challenges, teachers can start with simpler tasks and advance to more complex scenarios, mirroring mastery learning principles. Research has noted that such an approach provides appropriate scaffolding throughout the learning process and parallels how athletes or musicians incrementally build skills through practice and feedback (Ke et al., 2020; Craig, 2013; Huang et al., 2022).

**Cognitive Affective Model of Immersive Learning**

Contemporary theoretical frameworks specific to immersive learning have also been applied to avatar-based teacher training. One prominent model is the Cognitive Affective Model of Immersive Learning (CAMIL) proposed by Makransky and Petersen (2021). CAMIL suggests that the unique affordances of immersive VR—namely a strong sense of presence (being "there" in the virtual classroom) and agency (the ability to naturally interact with the environment)—influence learning through cognitive and affective pathways (Makransky and Petersen, 2021). In a teaching simulator, high presence might mean the trainee feels genuinely embedded in a classroom, and high agency means they can move and interact (e.g., write on the virtual board or physically approach a disruptive student). According to CAMIL, these features can increase a learner's interest, motivation, and self-efficacy while managing cognitive load, ultimately leading to better learning outcomes (Makransky and Petersen, 2021). Applying CAMIL to teacher training, we can infer that a fully immersive, interactive VR classroom (with responsive avatars) might produce greater engagement and skill uptake than a non-immersive role-play or video case study. This is supported by empirical comparisons: a recent study found that VR classroom simulations led to higher gains in teaching interest and self-efficacy for student teachers compared to traditional video-based training (Huang, 2023). The sense of realism provided by VR—including realistic student avatar appearance, behaviors, and even classroom visuals—contributes to what researchers call ecological validity, the authenticity of the experience (Docter et al., 2024). Docter et al. (2024) specifically stress ensuring the virtual classroom has authentic visuals and that avatar students exhibit natural postures, facial expressions, and gestures to mirror real student reactions. These realistic touches are informed by social learning theory; if trainees treat the avatars as real students, they are more likely to take the exercise seriously and develop meaningful skills (akin to Bandura's concept that effective practice requires realistic social context).

**Cognitive Load Theory**

Building on the cognitive factors highlighted by CAMIL, Cognitive Load Theory (CLT) offers further insight into how avatar realism might impact learning, particularly for novices (Sweller et al., 1998). CLT suggests that learning is hampered when the total cognitive load exceeds the learner's limited working memory capacity. Designing instruction to minimize extraneous cognitive load (load not essential for learning) is therefore crucial (Van Merriënboer & Sweller, 2005). In the context of VR simulations, high avatar realism (complex visuals and nuanced/unpredictable AI behaviors) could potentially impose high extraneous cognitive load. Novices might struggle to simultaneously manage the core teaching task (intrinsic load) while interpreting subtle or complex avatar cues (extraneous load), hindering schema construction (germane load) (Plass et al., 2010).

## Scaffolding

Conversely, lower avatar realism (e.g., stylized avatars with clearer, perhaps exaggerated cues) might initially be more beneficial by reducing extraneous cognitive load, allowing novices to focus resources on the teaching task itself (Plass et al., 2010). This aligns with the principle of scaffolding within Vygotsky's (1978) ZPD; managing cognitive load effectively keeps the learner within their optimal learning zone (Van Merriënboer & Sweller, 2005). Therefore, a scaffolded, graduated realism approach is theoretically supported: starting novices with less cognitively demanding (less realistic) avatars and gradually increasing fidelity and behavioral complexity as their expertise develops (cf. Craig, 2013; Dieker et al., 2014). This aims to provide a smoother learning path by managing cognitive demands effectively. This theoretical stance is supported by recent empirical findings. For instance, Westin & Öberg (2024) found that students who were aware they were interacting with an AI avatar performed worse on memory retention tasks, suggesting that this awareness divided their attention and increased extraneous cognitive load as they focused on the avatar's authenticity rather than the educational content.

## Four-Component Instructional Design

This concept of scaffolding is also central to other instructional design frameworks for complex learning, such as the Four-Component Instructional Design (4C/ID) model (Ross & Carney, 2017). The 4C/ID model structures training around authentic, whole-task practice, moving from simple to complex, with decreasing support (Van Merriënboer, 2002). A recent study in nursing education by Nasrollahi et al. (2025) applied this model and found that while high-fidelity (on-the-job) training led to the best performance, moderate-fidelity (simulation) was superior for learner satisfaction compared to high-fidelity environments. This finding from a parallel field reinforces the pedagogical hypothesis that an intermediate, controlled level of realism can be optimal for managing cognitive load and learner affect, providing further support for a graduated realism approach.

## Interpersonal Behavior Theory

Beyond managing cognitive load, the specific programming of AI avatars allows for embodying other theories directly related to interaction. Another conceptual model evident in student avatar research is the teacher interpersonal behavior theory. Docter et al. (2024) draw on an interpersonal behavior model (originally based on Leary's work) that maps teacher actions on axes of influence (dominance vs. submission) and proximity (cooperation vs. opposition). Novice teachers often struggle to find the right balance—for instance, they may be too passive or too accommodating (Docter et al., 2024). By embedding this model into the VR simulation's design, the AI avatars can be programmed to respond in ways that make the trainee aware of their interpersonal style. For example, an overly submissive approach by the trainee might cause some avatars to become unruly (since low dominance can invite off-task behavior), thus cueing the teacher to take more leadership in the classroom. In contrast, a domineering approach might provoke oppositional reactions from certain avatars, teaching the importance of warmth and cooperation. The use of such theory-driven behavior rules allows the simulation to serve as a conceptual learning tool: teachers don't just practice routines, they also implicitly learn theory (like interpersonal dynamics) by seeing it play out in real time. The avatars essentially embody educational theories, giving trainees concrete examples of abstract concepts. For future research, we can even imagine inputting specific educational theories into the AI avatars' programming so they exhibit behaviors that challenge the teacher in line with those theories.

In summary, VR teacher simulators are underpinned by experiential learning principles (Kolb, 1984), constructivist and socio-cultural theories (Vygotsky, 1978), considerations of cognitive load (Sweller et al., 1998), and emerging immersive learning models like CAMIL (Makransky & Petersen, 2021). Specific interaction models, such as the teacher interpersonal behavior theory, can be directly embodied by AI avatars (Docter et al., 2024). These simulators offer a unique convergence of theory and practice: trainees apply educational concepts in a realistic setting, which reinforces their conceptual understanding. By comparing and applying different theoretical lenses, researchers strive to optimize these simulations for maximum learning impact. However, leveraging these theories effectively also requires evaluating where current implementations, particularly regarding the optimal application of avatar realism and AI behavior modeling, fall short, which leads to the next section on research gaps.

**The Pedagogical Case for Graduated Realism**

Central to this review is the pedagogical hypothesis of graduated realism. This approach proposes that novice teachers, particularly those new to VR simulation training, should initially engage with environments featuring lower levels of avatar complexity, both visually and behaviorally. Instead of immediately confronting trainees with hyper-realistic avatars like those envisioned by TaoAvatar (Chen et al., 2025), training could commence with stylized, perhaps cartoonish or animated, avatars. These initial avatars might exhibit clearer, potentially exaggerated, facial expressions and body language, making non-verbal cues easier for novices to interpret. Concurrently, the simulated classroom scenarios presented at this early stage would involve less complex management challenges, posing a lighter cognitive and emotional load.

The rationale for this graduated approach is grounded in established learning principles, primarily managing cognitive load (Sweller et al., 1998; Plass et al., 2010) and mitigating performance anxiety (McGarr, 2020). High-fidelity avatars presenting nuanced, unpredictable behaviors within complex scenarios could impose excessive extraneous cognitive load on novices who are still mastering fundamental teaching skills. Conversely, lower-fidelity representations can reduce this load, potentially lowering anxiety and allowing trainees to focus cognitive resources on core pedagogical techniques (e.g., basic questioning, giving instructions) before needing to interpret subtle social cues or manage complex disruptions. This progression mirrors effective design principles often seen in complex skill acquisition domains, such as video games, where initial levels provide controlled challenges before difficulty gradually increases.

Recent research provides strong support for this tailored approach. Tan et al. (2025), in a study comparing text-only, deepfake, and mascot avatars, found that student preferences varied significantly based on learning habits and context. Task-oriented learners often preferred the simple, distraction-free text interface for quick questions, while engagement-oriented learners favored the more interactive avatars (Tan et al. 2025). This suggests that a one-size-fits-all approach to realism is suboptimal and that a graduated or context-dependent model may be more effective. This finding is echoed in other high-stakes professional training. For instance, a study in nurse practitioner education found that while both fully immersive and screen-based AI simulations were valuable, learners expressed a preference for the screen-based version due to greater ease of use and better support for group collaboration (Anthamatten & Holt, 2024). This further suggests that the optimal level of realism is a function of both the learner's needs and the specific learning activity.

Conceptually, graduated realism functions as a form of scaffolding (Vygotsky, 1978), dynamically adjusting the simulation's demands to keep the trainee operating within their Zone of Proximal Development (ZPD). As trainees develop competence and confidence through practice and reflection—potentially aided by session recordings and targeted feedback (perhaps even AI-driven analysis of performance within each stage)—the simulation would introduce progressively more realistic avatars (visually and behaviorally) and more complex scenarios. This ensures a tailored learning trajectory, culminating in practice within high-fidelity simulations that demand nuanced understanding and response, representing the peak challenge. Ideally, AI within the simulator could further personalize this progression based on individual performance, institutional curriculum goals, or learner objectives.

Despite the strong theoretical appeal of this approach, grounded in cognitive load theory and principles of scaffolding, there appears to be limited to no empirical research directly investigating the pedagogical effectiveness of graduated avatar realism in VR teacher training. Existing studies (as reviewed earlier) tend to evaluate the impact of a single type of simulator or a fixed level of realism, rather than systematically comparing different fidelities or testing the efficacy of a progressive, scaffolded model based on realism levels. This represents a significant gap in the current literature.

**Graduated Realism Visualization**

Stage 1: Animated/Cartoonish, Exaggerated facial expressions, easy teaching situation (Pixiv Inc, (n.d.))

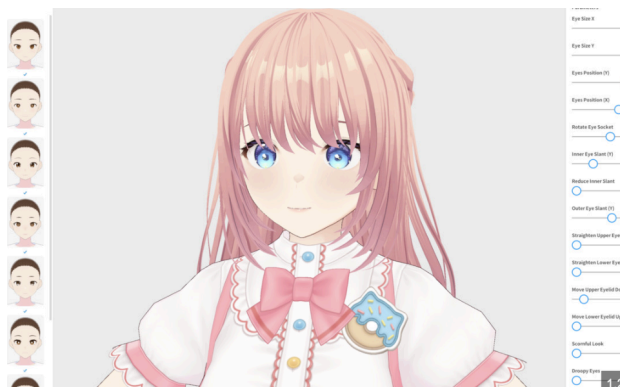

Stage 2: Slightly Realistic, Regular facial expressions and body language, medium teaching situations (Player Me, (n.d.))

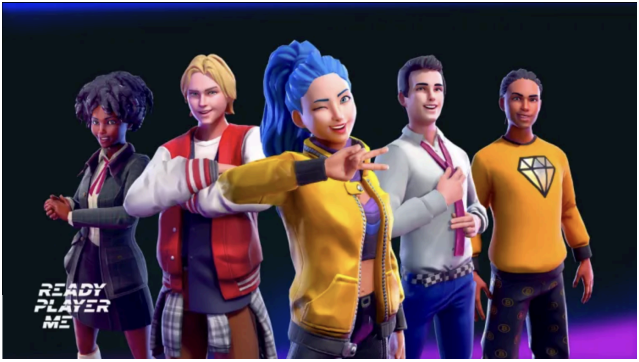

Stage 3: Hyper PhotoRealistic, subtle facial expressions and body language, advanced teaching situations possible (Chen et al., 2025)

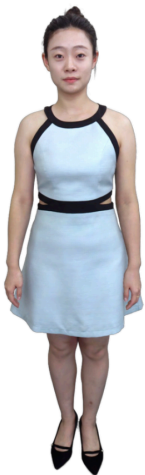

Relighting

## Ethical Considerations of AI in Teacher Simulations

The integration of AI, particularly large language models (LLMs), into educational simulations raises several critical ethical questions that researchers and developers must carefully address. Key themes involve the use of real student behavioral data for training, the potential for bias, and upholding professional values within these simulated interactions.

### Privacy and Data Protection

Training an AI on real student data (e.g., classroom transcripts, interaction records) involves highly sensitive information from minors, protected by laws like FERPA in the U.S. and ethical standards requiring consent and confidentiality. Thorough anonymization is necessary, yet even then, models might inadvertently mimic specific students or reveal private details, with the risk of LLMs "leaking" training data fragments in conversation (Fink et al., 2024). This raises complex issues of consent and fairness: have students and parents agreed to this use, and who owns the data integrated into the AI's knowledge?

Furthermore, AI educational services often utilize cloud infrastructures, potentially transmitting and storing sensitive data like voice recordings on external servers, which could be misused if breached (Fink et al., 2024). Encouraging natural conversation with AI avatars also carries the risk of users oversharing personal information, exacerbated by the documented tendency for users to form relationships with AI agents (Fink et al., 2024). Initiatives like the recently launched Center for Education Data Science Institute (EDSI) at the University of Maryland aim to facilitate access to high-quality educational data (EDSI, 2025), but such resources do not negate the fundamental ethical requirements. Any use of real student data necessitates strict data governance, robust anonymization, explicit consent, and secure infrastructure. Ideally, models should be trained primarily on synthetic or appropriately licensed public data whenever possible. Additionally, supervision by professional educators is crucial when AI avatars are used in critical settings involving minors (Fink et al., 2024), including ensuring trainee session recordings are secure and used appropriately for reflection or research with consent.

**Bias and Representativeness**

An education-specific LLM trained on data from limited demographics (e.g., predominantly suburban middle-class schools) may introduce significant bias, failing to simulate diverse student behaviors accurately and potentially providing a skewed training experience. Achieving representative AI avatars requires a wide range of diverse student data, yet this goal of "ultimate customization" (Fink et al., 2024) directly conflicts with privacy concerns and the ethical risks of using extensive real data.

A potential compromise involves using composite or synthetic data, creating realistic fictional profiles based on expert knowledge rather than raw transcripts (Fink et al., 2024). However, even with curated data, continuous auditing for bias is essential. LLMs inherit biases from their vast internet training data, which may contain harmful stereotypes or "inappropriate values" (Fink et al., 2024). Bias can manifest subtly in simulations, such as avatars responding differently based on the perceived gender of the teacher avatar or portraying student groups stereotypically. It is ethically imperative to align AI avatars with inclusive educational values. This necessitates careful technical approaches, such as using Retrieval-Augmented Generation (RAG) with vetted knowledge bases and/or fine-tuning LLMs on data reflecting positive norms, alongside rigorous testing to detect and mitigate bias (Fink et al., 2024). Transparency regarding these measures is vital for building teacher trust.

**Professional Values, Authenticity, and Simulating Challenging Behaviors**

A significant concern is that AI avatars, as machines, lack the inherent professional and ethical judgment of human educators. Teachers cultivate empathy, patience, and ethical standards over years; an AI lacking these might respond unproductively to trainee errors, such as reacting negatively to an insensitive comment without providing constructive guidance, or reinforcing the mistake by failing to recognize it (Fink et al., 2024). This underscores the need for caution against avatars operating unsupervised in high-stakes training scenarios (Fink et al., 2024) and highlights the importance of a human feedback loop (e.g., instructor monitoring, AI mentor agents).

This limitation becomes particularly salient when considering behavioral authenticity. Realistic teacher training arguably requires practicing responses to difficult, sometimes unethical, student behaviors (e.g., defiance, disrespectful language, bullying). However, simulating these authentically poses an ethical and technical dilemma. Standard AI safety filters often prevent models from generating such negative content, leading to unrealistically resilient or overly polite avatars (Docter et al., 2024; Fink et al., 2024). This challenge of imperfect realism extends beyond social behavior to cognitive performance; LLMs trained as "helpful assistants" also struggle to replicate the plausible errors and varied cognitive levels of actual students, often generating answers that are too perfect or advanced to be authentic (Wu et al., 2025). Moreover, an AI lacks the human capacity to model the impact of such events or guide the trainee through them ethically. Therefore, safely and effectively simulating these challenging realities necessitates careful design, potentially leveraging the graduated realism approach to introduce such scenarios with appropriate scaffolding and support. Over-reliance on the simulator as a perfect reflection of reality must also be avoided; reflection sessions with human supervisors are crucial for interpreting the simulated experience appropriately and mitigating risks like forming "inadequate relationships" with the AI (Fink et al., 2024) or misjudging one's own skill level.

The use of hyper-realistic avatars becomes even more complex when the avatar is a 'deepfake' of a known individual, such as the course instructor. Tan et al. (2025) found that familiarity with the lecturer enhanced perceived trustworthiness and that students appreciated the consistent teaching style. However, this also raises unique ethical considerations regarding consent, representation, and the potential for misplaced trust if the AI were to provide incorrect information.

**Teacher Autonomy and Trust**

Introducing sophisticated AI into teacher training raises broader questions about teacher autonomy and trust. Will extensive training with AI lead novices to over-rely on AI recommendations versus their own judgment or human mentorship? Studies show teachers often hold misconceptions and fears about AI, such as job replacement or unfair decision-making (Nazaretsky et al., 2022). Building trust requires transparency in how AI avatars function and empowerment, ensuring trainees understand the simulator is a tool to support their judgment, not replace it. Professional development that demystifies AI can increase teachers' willingness to use these tools effectively (Nazaretsky et al., 2022). Clearly communicating that avatar behaviors stem from models and parameters—and that unexpected responses might be glitches, not trainee faults—is crucial. This technology is a simulation, a tool for practice, not an infallible assessor of worth.

**Ethical Use of AI with Students**

Finally, the ethical design of AI avatars for training carries direct implications for their potential future use with actual K-12 students (e.g., for tutoring or practice). Ensuring the AI provides accurate information is paramount; while LLMs have improved, guarding against hallucinations in educational contexts requires diligence, potentially through custom LLMs fine-tuned on vetted data (Hughes, 2023). Using commercial models introduces their inherent data policies and biases. While open-source models offer more control, they require expertise. Furthermore, equity concerns arise if only well-funded programs can afford advanced AI simulations, potentially widening disparities in teacher preparation quality. Ethical implementation demands exploration of open-source or lower-cost solutions to ensure broad access.

In conclusion**,** the infusion of AI into VR teacher training brings powerful possibilities but also a responsibility to uphold privacy, fairness, and educational integrity. Researchers and developers advocate for measures like partial human oversight in sensitive contexts, strict data protection protocols, and ongoing ethics reviews (Fink et al., 2024). By proactively addressing these multifaceted considerations, we can work towards ensuring that AI-driven teacher simulators remain a force for good–enhancing teacher learning while respecting the values at the heart of education.

### Benefits and Future Directions

The convergence of Virtual Reality (VR) and Artificial Intelligence (AI) technologies within teacher education presents significant potential benefits for both pre-service and in-service teachers, opening exciting avenues for pedagogical innovation. This section summarizes the key advantages identified in the literature and discusses how evolving technologies might further enhance teacher training.

### Enhanced Practice Opportunities in a Controlled Environment

A primary benefit of VR simulations is the provision of a risk-free, controlled environment for deliberate practice. Unlike traditional field experiences, which offer limited and often high-stakes classroom time, VR allows trainees to accumulate significantly more "teaching hours" through repeated engagement with virtual lessons (McGarr, 2020). This repetition fosters competence and builds confidence. Akin to flight simulators preparing pilots for emergencies, teaching simulators enable novices to encounter and practice responding to diverse classroom challenges (e.g., disruptions, technology failures) that may arise infrequently during standard student teaching. This accelerated exposure can shorten the learning curve. While empirical validation is needed, extensive simulator training could potentially reduce the time required to develop qualified, classroom-ready teachers, contributing to addressing teacher shortages.

### Bridging Theory and Practice through Experiential Learning

As discussed in the theoretical framework section, VR simulations facilitate the application and testing of educational theories within concrete contexts, thereby deepening conceptual understanding (Dieker et al., 2014; McGarr, 2020). Teacher candidates can actively experiment with pedagogical strategies learned in coursework (e.g., specific classroom management techniques, differentiated instruction) and receive immediate, observable feedback through the AI avatars' responses. This experiential learning cycle enhances retention and the ability to transfer theoretical knowledge into effective practice–addressing a frequently cited weakness of traditional teacher preparation (Grossman et al., 2009; Huang et al., 2023). Studies suggest VR-trained teachers report higher self-efficacy, which is crucial for reducing early-career attrition (Huang et al., 2023). Moreover, recent evidence suggests these benefits extend to long-term knowledge retention. Li et al. (2024) found that while both video-based and immersive

VR training improved immediate competency, the IVR group demonstrated significantly better performance on delayed tests, suggesting the experiential nature of the simulation enhances long-term learning.

**Personalized, Scalable, and Varied Learning Experiences with AI**

The integration of AI-driven avatars significantly enhances the potential for individualized and varied learning. Practice scenarios can be tailored to specific trainee needs; for example, AI avatars can be programmed to frequently require clarification to help a trainee practice questioning techniques, or present persistent challenges to develop classroom authority skills (Docter et al., 2024). AI can also enable scalable, personalized feedback mechanisms, such as an AI coach providing reports on specific interaction patterns, offering a level of individualized attention difficult to achieve solely with human mentors (Fink et al., 2024). Furthermore, AI vastly expands the variety of scenarios that can be simulated feasibly and on-demand. This includes practicing responses to rare but critical incidents (e.g., bullying, emergency procedures) or exploring the use of new curricula and classroom configurations, benefiting both pre-service and in-service teachers seeking professional development.

**Fostering Reflection and Growth in a Safe Space**

The absence of real students creates a psychologically safe space where trainees can experiment, make mistakes, and even fail without fear of negative consequences for pupils or formal evaluation pressures (McGarr, 2020). This encourages a growth mindset, innovation, and risk-taking. Shy or anxious teachers can build confidence before facing live classrooms. Similar to how athletes use visualization, teachers can use VR rehearsals for mental and pedagogical preparation, reducing performance anxiety. The ability to record and review simulation sessions (perhaps with peers or mentors) further supports deep reflection on practice (Ke & Xu, 2020).

**Supporting Continuous Professional Development**

For practicing educators, AI-driven VR simulators offer accessible, on-demand coaching and professional development. Schools could establish VR labs for teachers to periodically engage with simulations, sharpening skills or practicing new techniques relevant to current classroom challenges. Aggregate, anonymized data from these sessions could also inform administrators about common professional development needs, allowing for more targeted support (while respecting privacy considerations).

**Enabling Global and Cross-Cultural Training**

VR simulations transcend geographical limitations, offering unique opportunities for cross-cultural teacher training. Educators could practice teaching in simulated classrooms reflecting different cultural norms or linguistic diversity, preparing them for international posts or increasingly multicultural domestic classrooms. AI language capabilities could enable avatars to speak various languages or dialects, providing valuable practice for teachers working in multilingual settings.

**Future Directions**

**Enhancing Realism and Functionality**

Future technological advancements promise to amplify these benefits further. The integration of multimodal AI, enabling avatars to process and respond to visual cues like gestures and potentially infer emotional states (Fink et al., 2024; Wang et al., 2024), will greatly enrich non-verbal communication practice and interactional realism, though challenges in processing speed and cost remain (Wang et al., 2024). Advances in haptic feedback and VR hardware could allow for more immersive physical interactions, such as handling virtual objects. Real-time AI coaching within the simulation, offering immediate feedback or intervention, represents another potential enhancement.

On the AI development side, creating dedicated educational LLMs, potentially trained on high-quality, ethically sourced data (perhaps via initiatives like EDSI, 2025), could yield safer, more aligned, and highly believable student avatars. Such models might even simulate student learning progression within a session, challenging teachers to adapt instruction in real-time based on formative assessment data gathered within the simulation itself, thus creating a comprehensive practice environment for all facets of teaching.

Exploring multi-user simulations allowing for co-teaching practice or peer observation/feedback offers another avenue. Simulating larger-scale scenarios like parent-teacher conferences could also broaden the training scope. Finally, ensuring accessibility remains paramount as the technology matures. Leveraging open-source platforms (Fink et al., 2024) and designing for users with varying needs (e.g., addressing motion sickness) will be crucial for equitable adoption.

**A Call for Computationally Efficient, Authentically Embodied, and Educationally Aligned Simulators**

The research landscape explored in this review reveals a clear trajectory for the next generation of AI-driven VR teaching simulators. While current models demonstrate remarkable capabilities, their limitations in computational efficiency, behavioral authenticity, and pedagogical alignment present significant barriers to widespread, effective implementation. To move from promising prototypes to scalable, transformative tools, the field must address three critical priorities.

**The Need for a Dedicated, Fine-Tuned Educational LLM**

The core challenge underlying many of the issues discussed—from unrealistic responses to ethical concerns about data privacy—is that current simulators rely on general-purpose LLMs that are not optimized for the educational context. The single most impactful advancement would be the development of a specialized LLM fine-tuned exclusively on high-quality, ethically sourced educational data, such as that being compiled by initiatives like EDSI (2025).

A recent breakthrough in conversational AI outside of education, from the "Sesame AI" project, provides a compelling technical blueprint (Schalkwyk et al., 2025). Their Conversational Speech Model (CSM) focuses on achieving "voice presence" by training a model to understand and generate not just text, but the paralinguistic cues of human speech—tone, rhythm, and emotion. They demonstrate that while general TTS models have achieved near-human naturalness in isolated speech, they fail on "prosodic appropriateness" within a conversational context (Schalkwyk et al., 2025). This is a direct parallel to the challenge in our field: general LLMs can generate correct answers but fail to capture the authentic performance of a student. An "Education-LLM" built on similar principles—trained on student data to replicate not just what students say, but how they say it, their hesitations, their common misconceptions, and their emotional states—and potentially built with more efficient architectures like the split-transformer design used in the CSM model (Schalkwyk et al., 2025)**,** would represent a monumental leap forward in achieving true behavioral authenticity.

**A Framework for Authentic, Probabilistic Student Personality: "Crazy Slots"**

Even with a fine-tuned Education-LLM, the challenge of generating varied, dynamic, and unique student behaviors remains. Multi-stage, chain-of-thought reasoning can produce plausible imperfections but at a prohibitive cost and latency (Wu et al., 2025; Tonga et al., 2025). This makes real-time, interactive simulation at scale currently unfeasible. The ideal framework must therefore be computationally efficient, operating in a single pass while still capturing the complex, unpredictable nature of student personality.

A promising path forward is a single-call framework inspired by probabilistic models and chance-based mechanics, a concept here termed "Crazy Slots". This architecture inverts the current paradigm: instead of tasking a large, slow LLM with reasoning its way to an imperfect answer, it uses a fast, lightweight probabilistic engine to instantly generate a detailed behavioral instruction, and then tasks the LLM to simply perform that instruction. This approach shifts the computational burden from sequential LLM-based logic to parallel retrieval and calculation, making real-time interaction possible.

At the core of this framework would be a Retrieval-Augmented Generation (RAG) database housing detailed and dynamic student profiles. These profiles would be far more than static knowledge graphs; they would be living, multi-layered models of each virtual student's identity. Drawing from the manifesto for this work, a profile for a student named "Devin", for example, would contain several integrated layers:

1. **A Cognitive Layer:** A knowledge graph representing his conceptual mastery. For instance, it would store the parameter that he has a 90% probability of correctly answering a question related to the 4x multiplication tables, but perhaps only a 60% probability for 7x tables. This layer directly addresses the need for plausible cognitive imperfection (Wu et al., 2025).
2. **An Affective Layer:** A probabilistic map of emotional states, directly corresponding to the "Ten Emotions" model. Each emotion (e.g., Confusion: 15%, Engagement: 85%, Joy: 95%) has a baseline probability, creating a unique emotional signature for the student.

3. **A Behavioral Layer:** This layer captures personality traits and contextual preferences that dictate interaction styles. It would store information such as Devin's openness to feedback, his social connections to other students in the virtual classroom, and his personal interests (e.g., enjoys Fortnite). This layer is critical for enabling culturally relevant teaching analogies and understanding relational dynamics, a factor shown to be significant in user-avatar interactions (Tan et al., 2025).

4. **A Dynamic Modifier Layer:** Crucially, these parameters are not static. This layer contains a set of rules that allow the student's profile to evolve in real-time based on the teacher's actions. For example, a compliment from the trainee might trigger a rule that temporarily increases Devin's pride/accomplishment parameter by 10% while decreasing his anxiety/shyness by 5%. Conversely, a harsh critique might lower his engagement and increase resentment. This creates a genuine, responsive feedback loop, making the avatar feel alive and reactive.

The "Crazy Slots" engine would activate during each interaction. When the trainee asks Devin a question, the engine performs a "spin"—it queries the RAG database to retrieve all relevant parameters from Devin's profile within the current instructional context. It then calculates a probabilistic outcome based on all these weighted layers, generating a high-level behavioral instruction. For instance, the instruction might be: [Action: Answer Correctly; Confidence: 85%; Emotion: Joy; Tone: Eager; Contextual_Note: Use Fortnite analogy if applicable]. This single, comprehensive instruction is then passed to the fine-tuned Education-LLM. The LLM's task is no longer to reason; its task is to perform—to "voice act" the specific persona and action described in the instruction, delivering the line, "It's 12! I got this!" with an enthusiastic tone.

This architecture directly addresses the limitations of current systems. It dramatically reduces latency and cost by replacing multi-step LLM chains with a single, targeted query. Furthermore, it provides a robust and elegant framework for implementing Graduated Realism not just visually, but behaviorally. A trainee could start with students whose parameters are simpler and more predictable, and as they gain skill, progress to interacting with students whose profiles contain more complex, volatile, and challenging emotional and cognitive parameters. By separating the student "soul" (the RAG database) from the "voice" (the LLM), this approach offers a path toward simulators that are not only more realistic but also, for the first time, truly scalable, potentially enabling the use of smaller, optimized models that can run on local hardware without constant reliance on expensive API calls.

**Discussion**

In conclusion, AI-driven VR teacher training simulators offer multifaceted benefits, providing experience, feedback, personalization, and flexibility often exceeding traditional methods. Continued advancements in AI and VR realism, coupled with careful research into their pedagogical impact (particularly concerning graduated realism and ethical implementation), can significantly enhance teacher preparation and ongoing professional development. While challenges remain, particularly regarding ethical design, data sourcing, and accessibility, the potential for these technologies to transform teacher learning—making hands-on practice and theory application a continuous, accessible, and richly informative process—is immense. This approach holds the promise of revolutionizing teacher education, potentially making simulator-based practice an integral component of certification and lifelong learning for educators.

The integration of AI-driven student avatars into Virtual Reality (VR) simulations marks a significant evolution in teacher education pedagogy. This review has traced the progression from early human-controlled virtual classrooms, like TeachLivE™, to sophisticated prototypes where generative AI models animate entire classes, as exemplified by Docter et al. (2024). These advancements, grounded in established conceptual learning theories—including constructivism, experiential learning (Kolb, 1984), scaffolding (Vygotsky, 1978), Cognitive Load Theory (Sweller et al., 1998), and immersive learning models like CAMIL (Makransky & Petersen, 2021)—offer previously unattainable opportunities for authentic, flexible, and individualized teacher practice. They hold immense potential for bridging the persistent gap between educational theory and classroom practice (McGarr, 2020; Huang et al., 2022).

Despite this promise, significant challenges and research gaps remain. Technical limitations persist, particularly in achieving nuanced behavioral authenticity, incorporating effective multimodal communication, ensuring avatar memory/continuity, and overcoming latency issues (Wang et al., 2024; Tonga et al., 2025). A prominent example of these limitations is observed in the multi-stage LLM-driven simulation framework by Wu et al. (2025), which, despite achieving high simulation accuracy for imperfect student behavior, noted significant computational cost and time inefficiency, requiring approximately 20 minutes per student across experimental settings. Further supporting this observation, Tonga et al. (2025), in their LLM-to-LLM tutoring simulation, explicitly cited 'computational constraints' as a reason for primarily experimenting with only a single hint iteration (N=1), indicating that even relatively simple multi-LLM interaction loops face significant practical barriers to scalability. Furthermore, there is a notable scarcity of empirical evidence rigorously evaluating the long-term effectiveness of these AI-driven systems compared to traditional methods or simpler simulations. Paramount among the challenges are the ethical considerations: ensuring student data privacy, mitigating algorithmic bias, achieving representative avatars, navigating the complexities of simulating challenging behaviors safely, and fostering appropriate teacher trust and autonomy (Fink et al., 2024; Nazaretsky et al., 2022). The pursuit of high-fidelity realism must be carefully balanced against these ethical imperatives and pedagogical needs.

Emerging from these theoretical considerations and practical challenges is the pedagogical hypothesis of graduated realism. This review suggests that, rather than aiming immediately for maximum fidelity, starting novice teachers with less complex, perhaps stylized avatars, and gradually increasing visual and behavioral realism as skills develop, may be a more effective approach grounded in scaffolding and cognitive load management principles. This specific pedagogical strategy, however, remains largely untested empirically, representing a critical area for future research.

Nevertheless, the potential benefits of AI-powered VR simulators are compelling. They offer enhanced opportunities for safe, repeatable practice, boosting teacher confidence and self-efficacy (Huang, 2023). They provide a unique venue for applying theory, receiving immediate feedback, and engaging in reflection (Ke & Xu, 2020). AI enables personalized, scalable training adaptable to individual needs and diverse scenarios, supporting both pre-service preparation and continuous professional development.

The field is poised for rapid advancement through ongoing improvements in VR hardware, AI capabilities (especially multimodal AI and dedicated educational LLMs), and pedagogical design. Guiding these developments responsibly requires cross-disciplinary collaboration among educators, AI scientists, ethicists, and designers. Future research should prioritize longitudinal studies on teacher effectiveness, comparative analyses of different realism levels (testing the graduated realism hypothesis), the development of ethical frameworks for AI avatar creation and data use, and strategies for ensuring equitable access, potentially through shared platforms or open-source initiatives (Fink et al., 2024).

**Conclusion**

In closing, AI-driven VR teacher training simulators represent a powerful and promising innovation. While not a solution, they offer a potent supplement to traditional methods. By enabling deliberate practice in increasingly realistic and responsive environments, they can help cultivate teacher skill and confidence. Realizing the vision of readily available virtual classrooms for practice and growth requires sustained research, open dialogue on ethical use, and continued investment. Addressing the identified gaps and challenges thoughtfully ensures these transformative technologies develop as a genuine force for good, ultimately contributing to a new generation of teachers better equipped for the complexities of 21st-century classrooms.

**Acknowledgements**

I would like to express my sincere gratitude to my advisors, Dr. Claudia Galindo and Dr. Fengfeng Ke, for their invaluable guidance and unwavering support throughout this research process. Their expertise was instrumental in shaping the direction and rigor of this review.

My gratitude also extends to the researchers and institutions who provided access to their technologies and shared their insights. I thank Dr. Docter and Huu Dat for the opportunity to test their generative AI prototype; Shannon Kane, Loren Jones, and Dan Levin for providing valuable observations on the Mursion platform; and the Faculty and students at the University of Johannesburg for access to the TeachLivE™ apparatus.

During the preparation of this manuscript, the author used large language models (LLMs) to assist with improving clarity, refining grammar, and providing structural feedback on drafts.

Finally, I wish to offer my deepest thanks to my family and friends for their endless patience and encouragement.

**Declarations**

**Conflict of Interest**
The author declares no conflict of interest.

**Funding**

This research did not receive any specific grant from funding agencies in the public, commercial, or not-for-profit sectors.